\documentclass[pre,amsmath,amssymb,superscriptaddress,showpacs,twocolumn]{revtex4-1}

\usepackage{amsmath,amssymb,graphicx}
\usepackage{enumitem}
\usepackage{hyperref}
\usepackage{amsbsy}
\usepackage{latexsym}
\usepackage{color}
\usepackage{graphicx}
\usepackage{psfrag}
\usepackage[normalem]{ulem}
\usepackage{bm}
\usepackage{lipsum}
\usepackage{color}
\usepackage{tikz}

\newcommand{\be}{\begin{equation}}
\newcommand{\ee}{\end{equation}}
\newcommand{\bea}{\begin{eqnarray}}
\newcommand{\eea}{\end{eqnarray}}

\newcommand{\comment}[1]{}

\begin{document} 

\title{Coarsening in Bent-core Liquid Crystals: Intermediate Splay Bend State en route to the Twist Bend Phase}

\author{Nishant Birdi}
\email{srz188382@sire.iitd.ac.in}
\affiliation{School of Interdisciplinary Research, Indian Institute of Technology, Hauz Khas, New Delhi 110016, India.}

\author{Nigel B. Wilding}
\email{nigel.wilding@bristol.ac.uk}
\affiliation{H.H. Wills Physics Laboratory, University of Bristol, Royal Fort, Bristol BS8 1TL, United Kingdom.}

\author{Sanjay Puri}
\email{purijnu@gmail.com}
\affiliation{School of Physical Sciences, Jawaharlal Nehru University, New Delhi 110067, India.}

\author{Varsha Banerjee}
\email{varsha@physics.iitd.ac.in}
\affiliation{School of Interdisciplinary Research, Indian Institute of Technology, Hauz Khas, New Delhi 110016, India.}
\affiliation{Department of Physics, Indian Institute of Technology, Hauz Khas, New Delhi 110016, India.}

\begin{abstract}
We use molecular dynamics simulations to study coarsening dynamics in achiral banana-shaped bent-core liquid crystals following a quench from the high concentration polar smectic (SmX) phase to lower concentrations that favor the exotic twist-bend (TB) phase. Our novel result is the identification of an intermediate splay-bend state emerging prior to the eventual TB phase. The latter coarsens via the annihilation of {\it beta lines} which are analogous to string defects in nematic liquid crystals. Our findings are relevant for a large class of chiral systems assembled from achiral entities. 
\end{abstract}

\maketitle

\begin{figure}[ht]
\includegraphics[width=0.45\textwidth]{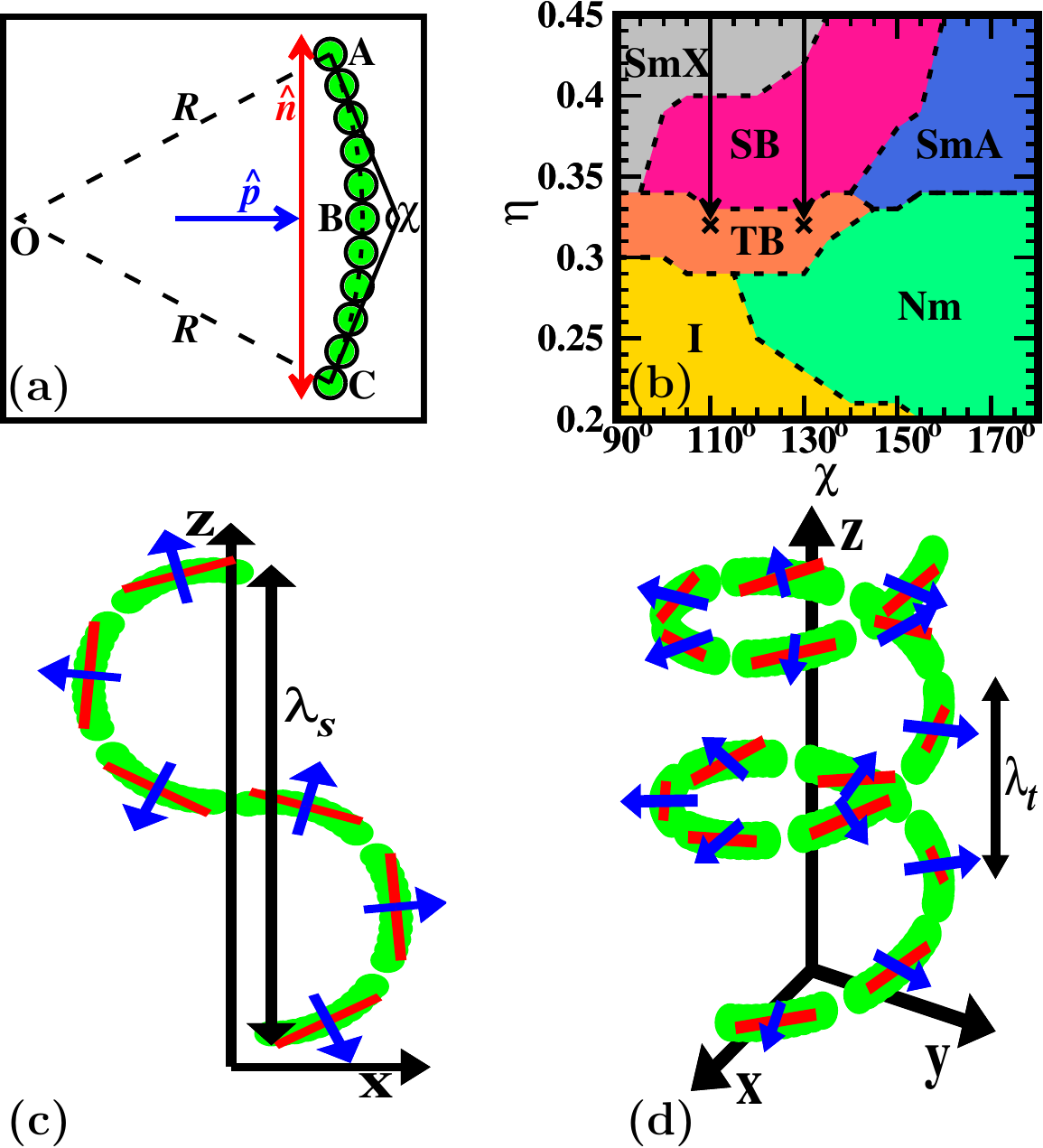}
\caption{(a) Schematic of a BCLC molecule with 11 beads, lying on an arc of a circle with radius $R$. It is characterized by the director $\hat{\boldsymbol{n}}$ (red) and the polarization $\hat{\boldsymbol{p}}$ (blue). (b) Phase diagram in the $(\chi,\eta)$-plane, figure adapted from \cite{Kubala_2022} (CC BY 4.0). The phases are as follows: isotropic (I), nematic (Nm), twist-bend (TB), splay-bend (SB), smectic A (SmA), and polar smectic (SmX). The arrows indicate the concentration quenches in our coarsening study. (c) SB structure with wavelength $\lambda_s$. (d) TB structure (left-handed helix) with pitch $\lambda_t$.}
\label{fig1}
\end{figure}

Liquid crystals (LCs) are an intermediate state of matter which combines fluidity with long-range orientational order \cite{Stephen_1974,deGennes_1995,Singh_2000,Priestly_2012,Andrienko_2018}. They are of huge technological interest for designing devices such as LC displays \cite{Chen_2018}, and are equally important in fundamental research for studies of topological defects \cite{Kleman_2008}. Rod-shaped molecules are the simplest example of LCs. These exhibit an orientationally ordered low-temperature uniaxial {\it nematic} (Nm) phase described by the {\it director} $\hat{\boldsymbol{n}}$. In the past two decades, there has been an intense excitement in this field due to investigations of banana-shaped {\it bent-core} liquid crystals (BCLCs). The alteration in shape produces novel phases with fundamentally different properties due to the emergence of {\it polar} order \cite{Lubensky_2002,Blanca_2005,Reddy_2006,Takezoe_2006,Etxebarria_2008,Jakli_2018} described by the {\it polarization} $\hat{\boldsymbol{p}}$ \cite{Meyer_1969} [see Fig.~(\ref{fig1})(a)].

In BCLCs, the uniform Nm phase can become unstable to form modulated nematics. In that case, there arises a  {\it bend} deformation of the $\hat{\boldsymbol{n}}$ field given by $\boldsymbol{B}=\hat{\boldsymbol{n}}\times (\nabla\times\hat{\boldsymbol{n}})$ \cite{Dozov_2001,Szmigielski_2023}. However, a constant $\boldsymbol{B}$ field cannot fill space uniformly. The geometrical frustration is then resolved either through formation of (a) a 3-$d$ chiral structure with a spontaneous {\it twist} $\mathcal{{T}}=\hat{\boldsymbol{n}}\cdot (\nabla\times\hat{\boldsymbol{n}})$ called the {\it twist-bend} (TB) phase; or (b) a 2-$d$ structure with {\it splay} distortions $\mathcal{S}=\hat{\boldsymbol{n}}(\nabla\cdot\hat{\boldsymbol{n}})$ termed as the {\it splay-bend} (SB) phase (see Fig.~\ref{fig1}). These structures were postulated by Meyer in 1973 \cite{Meyer_1976} and later by Dozov \cite{Dozov_2001}. The TB phase is characterized by $\boldsymbol{B},\mathcal{T}\ne 0$ and $\mathcal{S}=0$. If $\theta_0$ is the conical angle that the director of a BCLC molecule in the TB phase makes with the helical $z$-axis, then $\hat{\boldsymbol{n}}=[\cos(q_tz)\sin\theta_0,\sin(q_tz)\sin\theta_0,\cos\theta_0]$ and $\hat{\boldsymbol{p}}=[\sin(q_tz),-\cos(q_tz),0]$ \cite{Selinger_2013}. Here, $q_t=2\pi/\lambda_t$, where $\lambda_t$ is the {\it pitch} of the helix. The SB phase has $\boldsymbol{B},\mathcal{S} \ne 0$ and $\mathcal{T}=0$. The director of a BCLC molecule in the SB structure oscillates in a plane with a maximum angle $\theta^\prime$. We can then express $\hat{\boldsymbol{n}}=[\sin(\theta^\prime\sin(q_sz)),0,\cos(\theta^\prime\sin(q_sz))]$ and $\hat{\boldsymbol{p}}=[-\cos(\theta^\prime\sin(q_sz)),0,\sin(\theta^\prime\sin(q_sz))]$. Here, $q_s = 2\pi/\lambda_s$, where $\lambda_s$ is the SB wavelength \cite{Selinger_2013}. 

Experimental observations of the SB phase remain rare and are possible only under specific conditions \cite{Dijkastra_2020,Dijkastra_2021,Kotni_2022}. The TB phase was first observed a decade ago in the dimer CB7CB and is one of the most exotic and interesting LC phases \cite{Cestari_2011,Dong_2013}. The BCLC molecules are achiral but can organize into a (ambidextrous) {\it chiral} TB structure, providing a striking example of spontaneous mirror symmetry breaking in a liquid \cite{Fujii_2004,Kepa_2014,Yashima_2016,Tschierske_2018,Sang_2021}. This milestone discovery led to the rapid development of experimental synthesis methods \cite{Henderson_2011,Adlem_2013,Borshch_2013,Dong_2013,Mandle_2013,Sebastian_2014,Mandle_2014,Chen_2014,Mandle_2015,Gorecka_2015}, theoretical models \cite{Memmer_2002,Selinger_2013,Greco_2015,Chiappini_2019,Longa_2020,Dijkastra_2021,Kubala_2022,Szmigielski_2023} and explorations for technological applications \cite{Panov_2017,Panov_2021,Sreenilayam_2022,Antoni_2022,Khatun_2023,Meyer_2023}. Investigations in the past decade have primarily focused on understanding equilibrium properties. Non-equilibrium aspects have not been studied, the most fundamental being how SB and TB structures assemble from the BCLC molecules. 

In this letter, we address the above gap in the literature. We study the kinetics of phase transitions that occur when the system is rendered thermodynamically unstable by a sudden change of parameters \cite{Bray_2002,Puri_2004,Puri_2009}. The subsequent evolution is characterized by the emergence and growth of domains of the preferred phase. The domain growth depends on several factors such as symmetry of the order parameter, conservation laws, hydrodynamic flows, etc. The growth laws convey significant details of the ordering system, and the late-stage kinetics can be explained in terms of the defect dynamics. In this work, we use molecular dynamics (MD) simulations to perform deep quenches of a model BCLC system from a polar smectic  (SmX) phase to the TB phase by abruptly changing the concentration of BCLC molecules. This may be experimentally realized by adding/subtracting the solvent. We identify the nature of the defects and study their annihilation to understand the emergence of the chiral phase from achiral building blocks. Our most significant finding is the appearance of an intermediate SB phase that precedes the formation of the stable TB phase at late times. As the system evolves in configuration space so as to minimize its free energy, this suggests that the SB phase constitutes an easily accessible metastable state in the complex free energy landscape of the system.

Our MD simulations utilize a model of a BCLC molecule shown in Fig.~\ref{fig1}(a). It is constructed from eleven spherical beads (of mass $m$ and diameter $\sigma$) rigidly fixed with respect to one another, and evenly spaced along a circular arc with center $O$ and radius $R$  \cite{Greco_2015,Kubala_2022}. From the center of the terminating beads A and C, tangents are constructed which intersect at an angle $\chi$, an {\it inverse} measure of the structural bend. The two ordering directions for the BCLCs are $\hat{\boldsymbol{n}}$ and $\hat{\boldsymbol{p}}$. Beads on different molecules interact via the Weeks-Chandler-Andersen (WCA) potential \cite{Weeks_1971,Chandler_1983}. The emergence of different LC phases depends on the {\it bend angle} ($\chi$) and the {\it concentration} or packing fraction of beads ($\eta$). A rich phase diagram was presented earlier in \cite{Kubala_2022}. Fig.~\ref{fig1}(b) depicts an adapted schematic (CC BY 4.0). 
The schematics in Figs.~\ref{fig1}(c) and (d) show the arrangement of the BCLCs in the SB and TB structures, respectively.

The MD simulations are performed in the constant $NVT$ ensemble using LAMMPS \cite{Plimpton_1995,Lammps}. $N=9\times10^4$ molecules are simulated in a periodic cubic box of size $L_s\simeq117.443$ (all lengths are in units of $\sigma$). The equations of motion are integrated using the rigid body algorithm \cite{Kamberaj_2005}, and the system temperature $T=1$ is maintained constant via the Nos\'e-Hoover thermostat \cite{Martyna_1992}. The quenches indicated in Fig.~\ref{fig1}(b) are from the SmX phase ($\eta\simeq0.45$) to the TB phase ($\eta\simeq0.32$) for $\chi = 110^\circ$ and $130^\circ$. In the initial state $(t=0)$, the BCLC molecules are arranged on a lattice with $\hat{\boldsymbol{n}}$ parallel to the $z$ axis. The polarization $\hat{\boldsymbol{p}}$ of all molecules is co-aligned in a randomly selected direction. We then allow the system to evolve in time. All statistical measures have been averaged over 15 independent initial conditions with random choice of $\hat{\boldsymbol{p}}$.

\begin{figure}[ht]
\includegraphics[width=0.45\textwidth]{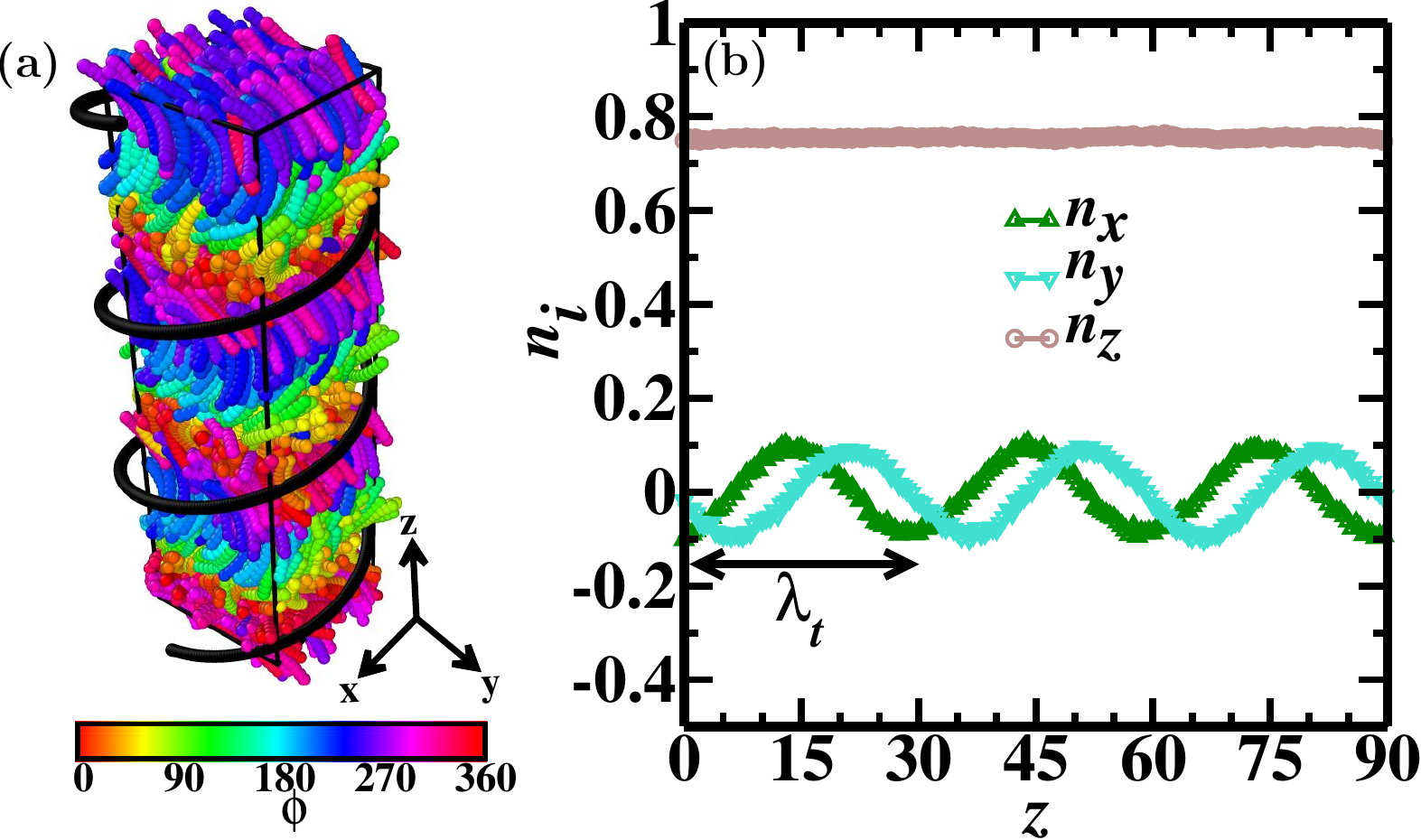}
\caption{(a) Equilibrated TB configuration in an elongated box for $\chi=110^\circ, \eta = 0.32$. The BCLCs are color-coded by $\phi$ -- the angle that the projection of $\hat{\boldsymbol{p}}$ on the $xy$-plane makes with the $x$-axis. (b) Corresponding variation of the laterally averaged components of $\hat{\boldsymbol{n}}$ along the helical $z$-axis.}
\label{fig2}
\end{figure}

We also perform a reference simulation with 3200 BCLC molecules in a box elongated along the $z$-axis ($L_x, L_y = 25.297, L_z=90$), setting it as the helical axis. It is relatively easy to drive this small system to the equilibrium TB phase. The resultant configuration is shown in Fig.~\ref{fig2}(a). We determine the TB properties: $\lambda_t$ from the period of the averaged $x$ and $y$ components of $\hat{\boldsymbol{n}}$; and $\theta_0$ from the corresponding $z$ component [see Fig.~\ref{fig2}(b)]. The values are: ($\lambda_t,\theta_0$) = (30,$39^\circ$) for $\chi=110^\circ$, and (45,$37^\circ$) for $\chi=130^\circ$. These values will be used later to evaluate the Ginzburg-Landau (GL) values for the magnitudes of $\mathcal{T}$ and ${\boldsymbol B}$, using the definitions mentioned earlier.

\begin{figure}[ht]
\includegraphics[width=0.45\textwidth]{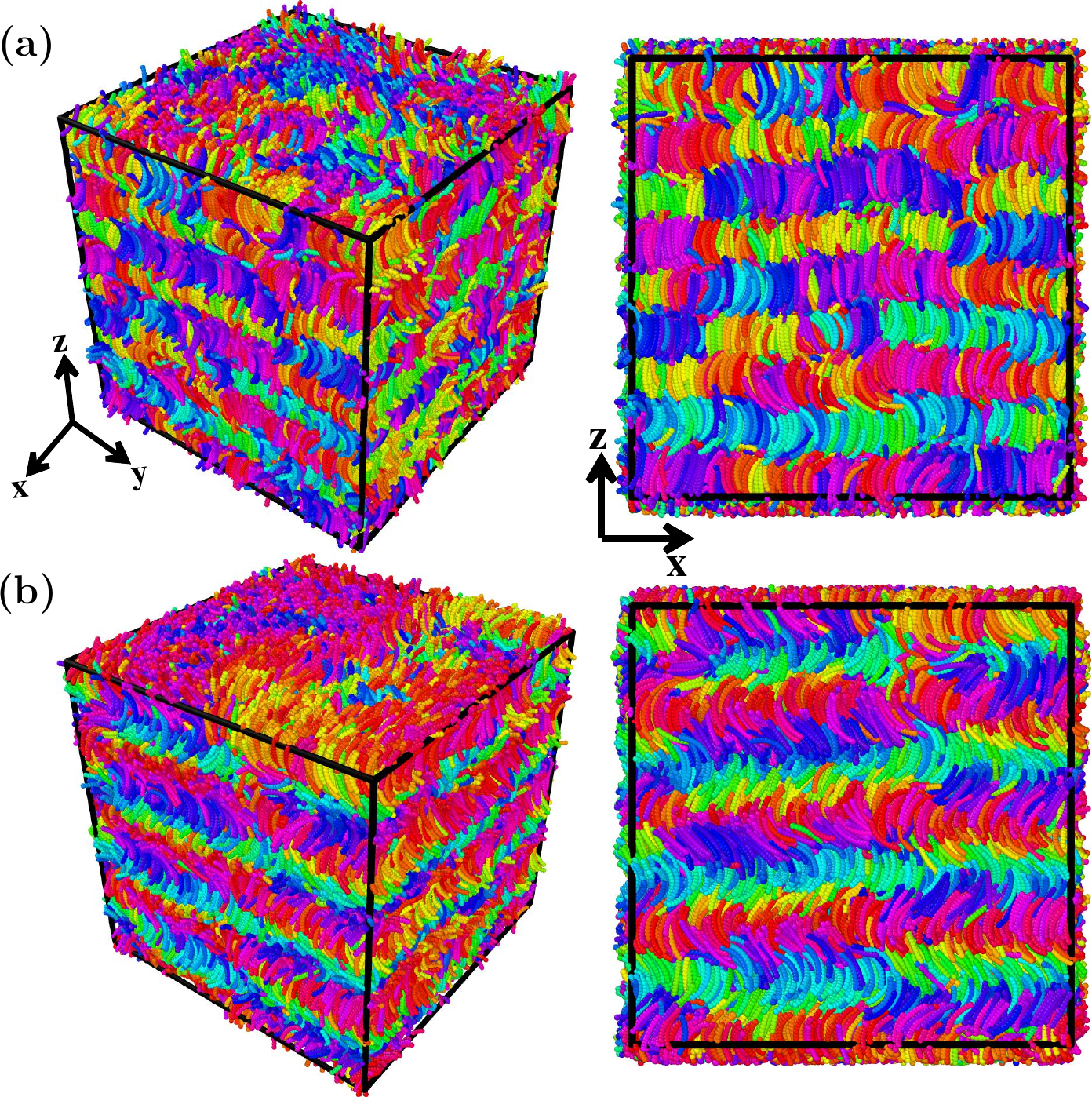}
\caption{Snapshots after a quench at $t=0$ from SmX $\rightarrow$ TB for $\chi=110^\circ, \eta=0.32$. The frames on the left show 3-$d$ snapshots at $t=$ (a) $2 \times 10^6$, and (b) $5 \times 10^6$. The frames on the right show the corresponding $(xz)$ cross-sections from $y=0$. The BCLCs are color-coded by $\phi$, as in Fig.~\ref{fig2}.}
\label{fig3}
\end{figure}

Now, let us show results from our domain growth studies, performed in a cubic box. In Fig.~\ref{fig3}, we present evolution snapshots for $\chi = 110^\circ$ at time $t$ (in units of MD time-steps) $=$ (a) $2\times10^6$ and (b) $5\times10^6$. The corresponding $xz$ cross-sections at $y=0$ are presented alongside. The BCLCs are color-coded as in Fig.~\ref{fig2}(a). The evolution demonstrates the development of helical order along the $z$-axis. In Fig.~\ref{fig3}(a), periodic modulations have developed along the $z$-axis. The layers with alternating colors shown in Fig.~\ref{fig3}(b) correspond to S-shaped SB structures. The late-time morphology in Fig.~\ref{fig3}(b) with continuously changing $\phi$ indicates the development of helical order characteristic of the TB structure [cf. Fig.~\ref{fig2}(a)].

To quantify the structural evolution, we evaluate the longitudinal (parallel) orientational correlations of $\boldsymbol{\hat{n}}$, $\boldsymbol{\hat{p}}$ and chirality $h$. The last of these captures the handedness of the helical organization of the BCLCs \cite{Stone_1978,Greco_2015,Kubala_2022}. These are defined as: (i) $C^{n}_{\parallel}(r_\parallel)=\sum_{i \neq j} c_{ij_{\parallel}}P_2(\boldsymbol{\hat{n}_i}\cdot\boldsymbol{\hat{n}_j})$; (ii) $C^{p}_{\parallel}(r_\parallel)=\sum_{i \neq j} c_{ij_{\parallel}} \boldsymbol{\hat{p}_i}\cdot\boldsymbol{\hat{p}_j}$; and (iii) $C^{h}_{\parallel}(r_\parallel)=\sum_{i \neq j} c_{ij_{\parallel}}  [(\boldsymbol{\hat{n}_i}\times\boldsymbol{\hat{n}_j})\cdot \boldsymbol{\hat{r}_{ij_{\parallel}}} ](\boldsymbol{\hat{n}_i}\cdot\boldsymbol{\hat{n}_j})$, respectively. Here, $r_\parallel$ is the intermolecular separation along the helical axis. Further, $c_{ij_{\parallel}}=\delta(r_{ij_{\parallel}}-r_\parallel)/ \sum_{i \neq j} \delta(r_{ij_{\parallel}}-r_\parallel)$, where $r_{ij_{\parallel}}$ is the separation between the $i$ and $j$ molecules along the helical axis. 

\begin{figure}[ht]
\includegraphics[width=0.45\textwidth]{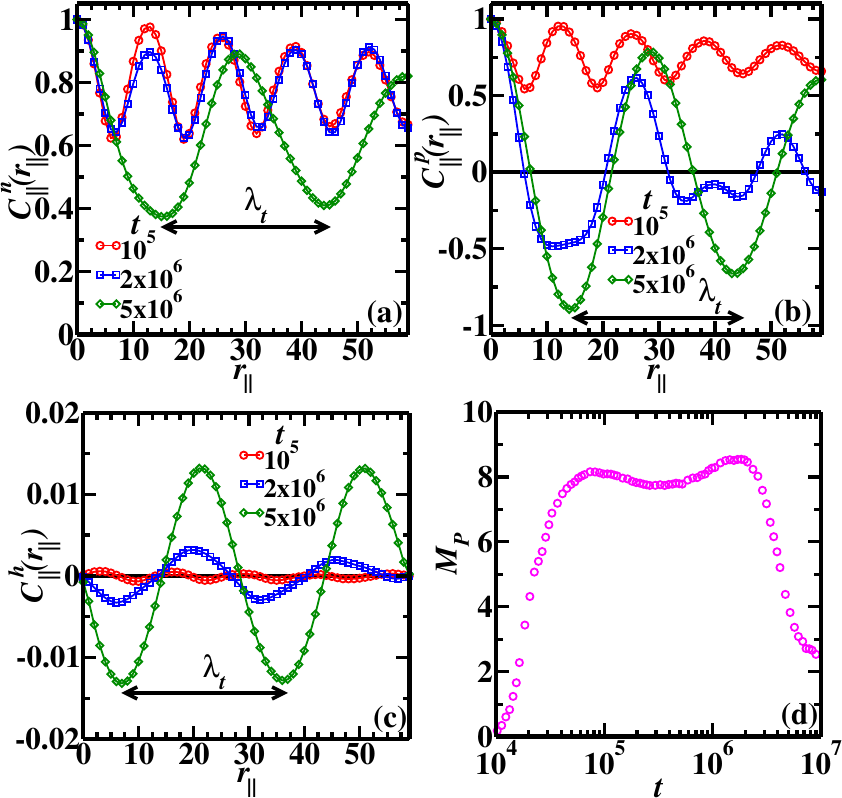}
\caption{Evolution of longitudinal correlation functions evaluated from: (a) $\hat{\boldsymbol{n}}$, (b) $\hat{\boldsymbol{p}}$, and (c) chirality $h$. We show data for $\chi=110^\circ, \eta=0.32$ at $t=10^5$, $2 \times 10^6$, and $5 \times 10^6$. (d) Corresponding time-dependence of $M_P$, defined in the text.}
\label{fig4}
\end{figure}

In Fig.~\ref{fig4}, we show (a) $C^{n}_{\parallel}(r_{\parallel})$ vs. $r_{\parallel}$, (b) $C^{p}_{\parallel}(r_{\parallel})$ vs. $r_{\parallel}$,  (c) $C^{h}_{\parallel}(r_{\parallel})$ vs. $r_{\parallel}$ for $\chi = 110^\circ$ at 3 $t$-values. Let us first focus on Figs.~\ref{fig4}(a)-(b). The initial condition ($t=0$) corresponds to the SmX phase with $C^{p}_{\parallel}(r_{\parallel}) = C^{n}_{\parallel}(r_{\parallel}) = 1$. At early times ($t=10^5$), both $C^{p}_{\parallel}$ and $C^{n}_{\parallel}$ develop oscillations about a non-zero average. For intermediate times ($t= 2 \times 10^6$), the oscillations grow and become characteristic of the SB phase (see Fig.~\ref{fig3}(a)) \cite{Kubala_2022}. Finally, for late times ($t= 5 \times 10^6$), the system has relaxed into the TB phase (see Fig.~\ref{fig3}(b)) with large-amplitude oscillations which are damped with distance. (In a perfect TB phase, the correlations do not decay with distance.) The periodicity of these oscillations ($\lambda_t \simeq 30$ in Figs.~\ref{fig3}(a)-(b)) determines the pitch of the TB phase. Notice that the $\hat{\boldsymbol{p}}$-field reflects helical order earlier than the $\hat{\boldsymbol{n}}$-field. We have also obtained similar data for $\chi=130^\circ$ (not shown here). The evolution to the TB phase is considerably faster in that case.

The correlation function $C^{h}_{\parallel}(r_{\parallel})$ vs. $r_{\parallel}$, which is a measure of chirality, is shown in Fig.~\ref{fig4}(c). In contrast to $C^{n}_{\parallel}$ and $C^{p}_{\parallel}$, it has been evaluated from a single morphology as the handedness of the emerging helix needs to be retained. The system is initially achiral since the $C^h_\parallel(r_{\parallel})$ correlations are negligible for $t<5\times10^6$. Periodic variations build up as time evolves, capturing the development of chirality that is characteristic of the TB phase. Note that the data in Fig.~\ref{fig4}(c) indicates a left-handed helix, and an opposite sign will be obtained for a right-handed helix.

To confirm the intermediate SB phase, we define an order parameter $M_P = \int |n_z(z)-\langle n_z(z) \rangle| dz$. From the definitions of $\hat{\boldsymbol{n}}$ for the SB and TB phases, $M_P$ should be positive in the SB phase and zero in the TB phase. Fig.~\ref{fig3}(d) shows $M_P$ vs. $t$ for $\chi = 110^\circ$. In accord with the results of Figs.~\ref{fig4}(a)-(c), $M_P$ indeed exhibits a sharp drop near $t=2\times10^6$, confirming that the SmX-TB kinetic transformation traverses an intermediate SB state. As mentioned earlier, this result is the principal finding of our study.

To better elucidate the coarsening behavior, it is crucial to understand the nature of defects whose annihilation gives rise to domain growth. In the TB phase, the dominant defects are the {\it beta lines} that correspond to regions where the bend field $\boldsymbol{B} \simeq 0$ \cite{Binysh_2020}. To identify them, we coarse-grain the evolution snapshots by dividing the simulation box into a simple cubic lattice of overlapping sub-boxes of size $(10)^3$ each. The center of each sub-box defines a lattice site with  a director $\hat{\boldsymbol{n}}$, which is an average over all BCLCs with centers in the sub-box. As the boxes are overlapping, the size of the coarse-grained lattice is also $(118)^3$. The advantage of coarse-graining is that it reduces fluctuations in the system, and is analogous to a numerical renormalization group procedure.

\begin{figure}[ht]
\includegraphics[width=0.45\textwidth]{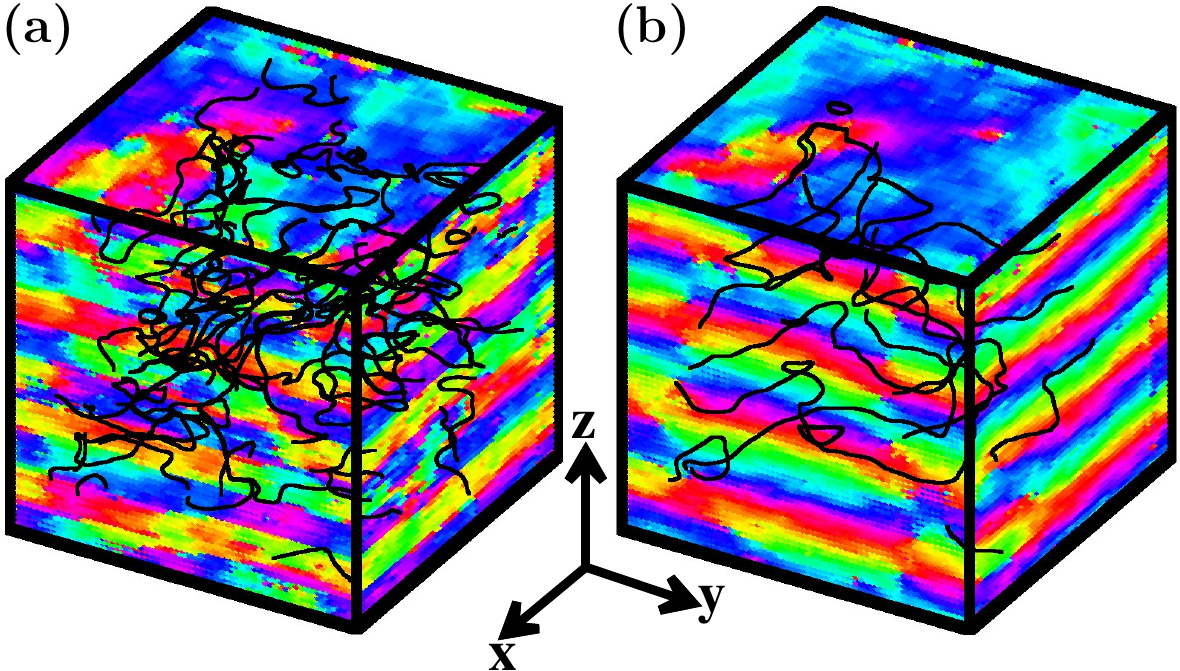}
\caption{Bend field for the snapshots in Fig.~\ref{fig3}. The frames show $\boldsymbol{B}$ (color-coded by $\phi$) at $t=$ (a) $2 \times 10^6$, and (b) $5 \times 10^6$. The {\it beta lines}, defect regions where $\boldsymbol{B} \simeq 0$, are marked black.}
\label{fig5}
\end{figure}

The derivatives of $\hat{\boldsymbol{n}}$, which are numerically approximated using the central-difference method with accuracy of order 8, are used to obtain the $\boldsymbol{B}$ and $\mathcal{T}$ fields at each lattice site. Figs.~\ref{fig5}(a)-(b) show the $\boldsymbol{B}$-field for the snapshots in Fig.~\ref{fig3}. The beta lines
are shown in black and their density decreases as coarsening proceeds. In the TB phase, $\boldsymbol{{B}}=q_t \sin \theta_0 \cos \theta_0 [-\sin(q_tz),\cos(q_tz),0]$. This anisotropic form results in modulations along the helical $z$-axis and uniformity in the planes perpendicular to it. Fig.~\ref{fig5}(b) captures this feature in the form of colored stripes that repeat in the $z$-direction. With time, the helical ordering grows and the beta lines annihilate. On the time-scales of our simulation, the beta lines do not exhibit marked anisotropy. They are reminiscent of vortex strings in 3-$d$ liquid crystals or the XY model.

What will be the signatures of these defects in coarsening experiments? Small-angle scattering experiments measure the structure factor $S(\vec{k},t)$, where $\vec{k}$ is the wave vector of the scattered beam \cite{Puri_2004,Puri_2009}.  If the ordering system is characterized by a single length scale $L(t)$, the structure factor obeys {\it dynamical scaling}: $S(k,t) = L^d f(kL)$, where $f(x)$ is a scaling function. The nature of the defects can be interpreted from the tail of the structure factor. Continuous $O(n)$ spin models exhibit the {\it generalized Porod law}, with the asymptotic form: $S(k,t) \sim k^{-(d+n)}$ \cite{Porod_1982,Yono_1988,Bray_1991}. For $n=1$, the defects are interfaces, and the corresponding scattering function exhibits the usual {\it Porod law}: $S(k,t) \sim k^{-(d+1)}$. For $n>1$, the different topological defects are vortices ($n=2,d=2$), strings ($n=2,d=3$) and monopoles or hedgehogs ($n=3,d=3$).

\begin{figure}[ht]
\includegraphics[width=0.45\textwidth]{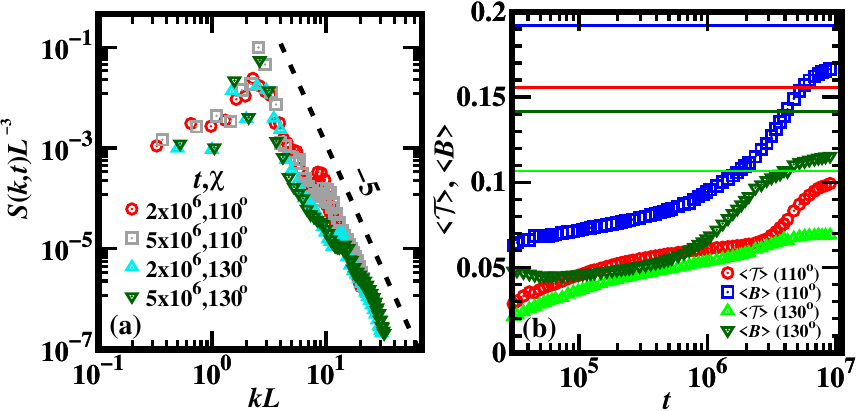}
\caption{(a) Log-log plot of the scaled structure factors of the $\boldsymbol{B}$-field for $\eta = 0.32$ and $\chi=110^\circ, 130^\circ$. The dashed line denotes the generalized Porod tail for vortex strings in $d=3$. (b) Time-dependence of $\langle \mathcal{T} \rangle$ and $\langle B \rangle$, where the angular brackets denote a spatial average. The horizontal lines show the corresponding GL values in the TB phase: $\mathcal{T}_{\rm GL}(110^\circ) \simeq 0.1555; \mathcal{T}_{\rm GL}(130^\circ) \simeq 0.1067; B_{\rm GL}(110^\circ) \simeq 0.1921; B_{\rm GL}(110^\circ) \simeq 0.1416$ \cite{Selinger_2013}.}
\label{fig6}
\end{figure}

As the coarsening in the $\boldsymbol{B}$ field is driven by isotropic vortex strings, we expect the corresponding structure factor to behave as $S(k,t) \sim k^{-5}$ at large $k$. In Fig.~\ref{fig6}(a), we show a scaling plot of the structure factor: $L^3 S(k,t)$ vs. $k L$ at 2 times for $\chi = 110^\circ, 130^\circ$. The length scale $L$ is defined as the inverse of the first moment of $S(k,t)$. The scaled data sets show a reasonable collapse, confirming dynamical scaling. Moreover, the structure factor tail is consistent with the expected Porod tail. For completeness, the spatial averages of $\mathcal{T}$ and $B$ (denoted as $\langle \mathcal{T} \rangle$ and $\langle B \rangle$) are shown in Fig.~\ref{fig6}(b). The horizontal lines denote the corresponding GL values of $\mathcal T$ and $B$. Not surprisingly, the GL mean-field values of these quantities \cite{Selinger_2013} overestimate our simulation results (which include fluctuations).

To conclude, we have performed extensive MD simulations in $d=3$ to study phase transition kinetics in BCLCs. This system is a liquid with achiral constituents, but exhibits remarkable phases. These include the achiral S-shaped planar SB phase and the (chiral) helical TB phase. We have focused on understanding the emergence of the exotic TB phase by monitoring the coarsening dynamics after a concentration quench from the SmX$\rightarrow$TB phase. Our unexpected finding is that this transition occurs through an intermediate SB structure.  The spontaneous formation of a metastable phase preceding the emergence of the stable phase during  spinodal decomposition has been observed in many experiments in the last decade. It is generic to the phase ordering kinetics associated with the recently introduced {\it landscape-inversion phase transitions} \cite{Alert_2014,Alert_2016}. We characterize the dynamic pathway SmX $\rightarrow$ SB $\rightarrow$ TB via appropriate correlation functions and structure factors. The dominant defects observed in the TB phase, the {\it beta lines}, are analogous to string defects in $d=3$ NmLCs \cite{Birdi_2020,Birdi_2022}. Their signature is a generalized Porod decay in the relevant structure factor: $S(k,t) \sim k^{-5}$.

We expect that the framework developed for studying BCLCs, along with our novel numerical results, will be important for a large class of systems with helical order. Experimental work on the chiral phases of BCLCs is limited. We hope that the current work will inspire and motivate further experimental studies of this system with its rich variety of equilibrium and non-equilibrium phenomena. \\
\ \\
\noindent{\bf Acknowledgments}: NB acknowledges UGC, India for a senior research fellowship. VB acknowledges partial financial support from SERB India via a research grant. NB and VB gratefully acknowledge the High Performance Computing (HPC) facility at IIT Delhi for computational resources.

\newpage
\bibliography{ref}

\end{document}